\documentclass[prl,twocolumn,nofootinbib, preprintnumbers, superscriptaddress]{revtex4}

\usepackage{amsmath, amssymb, slashed, braket, bm}
\usepackage{graphicx}
\usepackage{epstopdf}
\usepackage{float,appendix}
\usepackage[colorlinks=true,
            linkcolor=blue,
            urlcolor=blue,
            citecolor=green,          
            bookmarks=true,
            bookmarksnumbered=true,
            breaklinks=true,
            pdfpagemode=FullScreen,
            pdfstartview=FitBH]{hyperref}
\usepackage{esint}
\usepackage[normalem]{ulem}
\usepackage{siunitx}
\usepackage{multirow}

\usepackage{xcolor}

\definecolor{Orange}{cmyk}{0,0.61,0.87,0}
\definecolor{JungleGreen}{cmyk}{0.99,0,0.52,0}
\definecolor{OliveGreen}{cmyk}{0.64,0,0.95,0.40}
\definecolor{Brown}{cmyk}{0,0.81,1,0.60}
\definecolor{RoyalBlue}{cmyk}{0.71,0.53,0,0.12}
\definecolor{Gray}{cmyk}{0,0,0,0.40}
\definecolor{LightPink}{cmyk}{0.0,0.25,0,0}
\definecolor{LLightPink}{cmyk}{0.0,0.10,0,0}
\definecolor{LightBlue}{cmyk}{0.25,0,0,0}
\definecolor{LightGray}{cmyk}{0,0,0,0.2}

\definecolor{gesfpurple}{rgb}{0.47,0.19,0.42}

\definecolor{gesflanse}{rgb}{0.00,0.50,0.50}

\definecolor{gesfblue}{rgb}{0.08,0.42,0.76}

\definecolor{gesfred}{rgb}{1,0,0}

\definecolor{gesfwhite}{rgb}{1,1,1}

\definecolor{gesfblack}{rgb}{0,0,0}

\newcommand{\geqn}[1]{Eq.\,\hypersetup{linkcolor=blue}(\ref{#1})\hypersetup{linkcolor=blue}}
\newcommand{\gfig}[1]{{\hypersetup{linkcolor=violet}Fig.\,\ref{#1}\hypersetup{linkcolor=blue}}}

\newcommand\redout{\bgroup\markoverwith{\textcolor{red}{\rule[.5ex]{2pt}{0.4pt}}}\ULon}

\graphicspath{{Figures/}}

\begin{document}

\title{Torsion Balance Experiments Enable Direct Detection of Sub-eV Dark Matter}

\author{Shigeki Matsumoto} \email{shigeki.matsumoto@ipmu.jp}
\affiliation{Kavli IPMU (WPI), UTIAS, University of Tokyo, Kashiwa, 277-8583, Japan}

\author{Jie Sheng}
\email{jie.sheng@ipmu.jp}
\affiliation{Kavli IPMU (WPI), UTIAS, University of Tokyo, Kashiwa, 277-8583, Japan}
\affiliation{Tsung-Dao Lee Institute,  Shanghai Jiao Tong University, 201210, China}

\author{Chuan-Yang Xing}
\email{cyxing@upc.edu.cn}
\affiliation{College of Science, China University of Petroleum (East China), Qingdao 266580, China}

\author{Lin Zhu} 
\email{zhulin36@mail.hust.edu.cn}
    \affiliation{National Gravitation Laboratory, MOE Key Laboratory of Fundamental Physical Quantities Measurement \& Hubei Key Laboratory of Gravitation and Quantum Physics, School of Physics, Huazhong University of Science and Technology, Wuhan 430074, People's Republic of China}

\begin{abstract}

Light dark matter with sub-eV masses has a high number density in our galaxy, and its scattering cross section with macroscopic objects can be significantly enhanced by coherence effects. Repeated scattering with a target object can induce a measurable acceleration. Torsion balance experiments with geometric asymmetry are, in principle, capable of detecting such signals. Our analysis shows that existing torsion balances designed to test the Equivalence Principle already place the most stringent constraints on DM-nucleon scattering in the $(10^{-2}, 1)\,$eV mass range.
\end{abstract}

\maketitle

\section{Introduction} 

\noindent
Numerous astronomical and cosmological observations have confirmed the existence of dark matter (DM)~\cite{Young:2016ala}, which accounts for approximately 80\% of the total matter in the Universe~\cite{Planck:2018vyg}. While DM is widely believed to consist of elementary particles, its mass and interaction properties remain largely unknown~\cite{Bertone:2004pz, Arbey:2021gdg}.
Among possible interactions with the Standard Model (SM), elastic scattering is one of the most widely studied, and at low energies it is parametrized by an effective quadratic coupling of the form $\chi^{2}\mathcal{O}_{\rm SM}$, where the DM field $\chi$ couples bilinearly to an SM operator $\mathcal{O}_{\rm SM}$.
Since the proposal of DM--nucleus scattering as a means to detect weakly interacting massive particles (WIMPs)~\cite{Goodman:1984dc}, direct detection experiments targeting such elastic scattering have become a leading approach to probe DM nature~\cite{Lin:2019uvt, Cooley:2021rws}.

In the laboratory frame, DM particles have typical velocities of order $10^{-3}$ (in units of the speed of light), corresponding to kinetic energies around $10^{-6}$ times their mass. In elastic scattering processes, the energy transfer is always smaller than the incoming DM energy. As a result, for DM masses above the GeV scale, experimental efforts focus on detecting nuclear recoil signals with keV-scale thresholds~\cite{Billard:2021uyg, Akerib:2022ort}. For MeV--GeV DM, detection strategies shift toward eV-scale excitations in crystals or electron scattering~\cite{Essig:2011nj, Knapen:2017ekk, Griffin:2018bjn, Coskuner:2021qxo, Campbell-Deem:2022fqm, Mitridate:2022tnv, Ge:2025itf}. For even lighter (keV-MeV) DM, novel approaches employing ultra-low-threshold platforms, such as cold-atom interferometers or superconducting detectors, have been proposed~\cite{Afek:2021vjy, Essig:2022dfa, Baker:2023kwz}.

For DM masses below the keV scale, quadratic couplings arise in well-studied scenarios such as a $Z_2$-even complex scalar~\cite{Olive:2007aj, Hees:2018fpg, Bouley:2022eer, Gan:2025icr} or Higgs-portal models~\cite{Banerjee:2022sqg}. The observed abundance of such DM can be readily generated via the standard misalignment mechanism~\cite{Preskill:1982cy, Abbott:1982af, Dine:1982ah, Bouley:2022eer}.
However, in the mass regime below the keV scale, the DM kinetic energy is typically below $10^{-3}\,\mathrm{eV}$, where no established experimental technique is currently sensitive to such tiny energy transfers. If quadratically coupled DM is ultralight (i.e., below the meV scale), it can instead induce variations in fundamental constants, which can be probed by atomic clocks~\cite{Derevianko:2013oaa, Stadnik:2016zkf, Kalaydzhyan:2017jtv, Zhang:2022ewz}, interferometry~\cite{Stadnik:2014tta, Masia-Roig:2022net}, fifth-force experiments~\cite{Hees:2018fpg}, and astrophysical or cosmological observations~\cite{Olive:2007aj, Stadnik:2014cea, Stadnik:2015kia, Stadnik:2015uka, Blas:2016ddr, Blas:2019hxz, Dailey:2020sxa, Bouley:2022eer, Banerjee:2022sqg, Acevedo:2025rqu, Gan:2025icr}, among others~\cite{Stadnik:2022gaa}. Nevertheless, the sensitivity of these probes degrades rapidly for heavier DM because of the reduced number density. Detecting quadratically coupled DM with eV-scale mass remains an open challenge.


Although the energy transfer in eV-scale DM scattering is suppressed by the target mass, the momentum transfer need not be. Moreover, light (sub-eV) DM can exhibit greatly enhanced scattering rates with macroscopic targets due to its large number density and coherence effects~\cite{Fukuda:2018omk, Fukuda:2021drn, Luo:2024ocg}. Crucially, the cumulative effect of continuous scattering, namely, the net acceleration imparted to the target, provides a novel observable, shifting the focus away from traditional single-scatter energy deposition~\cite{Domcke:2017aqj, Graham:2015ifn, Carney:2019cio, Day:2023mkb}.
We note that experiments exploiting coherence to probe sub-eV DM, such as helioscopes~\cite{IAXO:2025ltd} and dielectric haloscopes~\cite{Jaeckel:2013eha, MADMAX:2024jnp, Chiles:2021gxk}, are sensitive to DM--photon conversion but not to elastic scattering.

In this work, we point out that asymmetries in the external size~\cite{Luo:2024ocg} and/or the internal structure of the test bodies in torsion-balance experiments can be exploited to detect DM-induced acceleration. We present the first systematic analysis of state-of-the-art torsion-balance experiments designed to test the Equivalence Principle (EP), and use them to constrain interactions between quadratically coupled light DM and nucleons. This yields robust, and currently the most stringent, direct-detection limits in the dark matter mass range $m_\chi\simeq (10^{-2},1)\,\mathrm{eV}$.

\section{Dark Matter-Induced Acceleration} 


\noindent
In the Milky Way, DM particle velocities follow an isotropic Boltzmann distribution. The Solar System orbits the Galactic Center at $v_S \sim 10^{-3}$\,\cite{Baxter:2021pqo}, creating relative motion with respect to the DM halo. From the Solar System's perspective, the DM fluid flows like a wind, referred to as the \textit{DM wind}. Since Earth's orbital speed around the Sun is much smaller, the DM wind velocity in the laboratory frame can be approximated as $v_\chi \simeq v_S \sim 10^{-3}$, directed opposite to the Solar motion.

Assuming DM interacts with nucleons ($N = n, p$) via a scattering cross-section $\sigma_{\chi N}$, it transfers momentum $q$ during scattering. For very light DM ($m_\chi \lesssim 1\,\mathrm{eV}$), the scattering is elastic, with $q \simeq m_\chi v_\chi$, comparable to the incident DM momentum. These interactions lead to DM scattering off macroscopic objects, characterized by a total cross-section $\sigma_{\mathrm{tot}}$. The resulting force from the DM wind, whose flux is $\rho_\chi v_\chi / m_\chi$, with $\rho_\chi \simeq 0.4\,\mathrm{GeV}/\mathrm{cm}^3$ the local DM density, is estimated as $F \sim \rho_\chi v_\chi \sigma_{\mathrm{tot}} q / m_\chi$. With $m_{\mathrm{tot}}$ being the total mass of a target, this force induces an acceleration on the target object as
\begin{equation}
    a \sim
    \frac{1}{m_{\mathrm{tot}}}
    \frac{\rho_\chi}{m_\chi}
    \sigma_{\mathrm{tot}} v_\chi q \simeq \frac{\rho_\chi \sigma_{\mathrm{tot}} v_\chi^2}{m_{\mathrm{tot}}}.
    \label{acc_chi_N}
\end{equation}

According to the uncertainty principle, the coherent scattering length is defined as $\lambda \equiv 1/q$, where $q \equiv |\mathbf{q}| \sim m_\chi v_\chi \simeq 10^{-3} m_\chi$. Within this coherent length, individual scattering events between the incident particle and constituents of the target become indistinguishable, requiring coherent summation of probability amplitudes. For instance, when $\lambda$ exceeds the nuclear radius, the DM-nucleus scattering cross-section is enhanced to $\sigma_{\chi A} = A^2 \sigma_{\chi N}$, where $A$ is the atomic mass number. This $A^2$ enhancement has already been experimentally confirmed in coherent neutrino-nucleus scattering\,\cite{COHERENT:2017ipa}.

When $m_\chi \lesssim 1\,\mathrm{eV}$, the coherent scattering length satisfies $\lambda \gtrsim 1\,\mu\mathrm{m}$, far exceeding typical interatomic distances. As a result, DM scattering off a macroscopic object must account for interference between different atomic sites, and the total cross-section becomes $\sigma_{\mathrm{tot}} = \sum_{i,j}^{N_A} e^{i \mathbf{q} \cdot \Delta \mathbf{r}_{ij}}\, \sigma_{\chi A}$\,\cite{Luo:2024ocg}. Here, the displacement between two nuclei is $\Delta \mathbf{r}_{ij} \equiv \mathbf{r}_i - \mathbf{r}_j$, typically of the order of the target size. When the coherence length exceeds the target size, i.e., $1/q \gg |\Delta \mathbf{r}_{ij}|$, the phase factor approaches unity: $e^{i \mathbf{q} \cdot \Delta \mathbf{r}_{ij}} \simeq 1$. In this regime, the total cross-section is enhanced by a factor of $N_A^2$, where $N_A$ denotes the number of atoms in the target. For instance, with $m_\chi = 10^{-2}\,\mathrm{eV}$ and $\sigma_{\chi A} = 10^{-50}\,\mathrm{cm}^2$, the total scattering rate for a macroscopic target with $N_A \sim 10^{24}$ reaches $\sim 10^{16}\,\mathrm{s}^{-1}$, resulting in a measurable acceleration,
\begin{equation}
    a \simeq 4 \cdot 10^{-12}\,\mathrm{cm/s^2} \times 
    \left( \frac{\sigma_{\chi A}}{10^{-50}\,\mathrm{cm^2}} \right)
    \left( \frac{N_A}{10^{24}} \right) .
\end{equation}
This shows that the DM wind continuously exerts a small force on macroscopic objects, including ourselves.

For a general momentum transfer $\mathbf{q}$, the sum of phase factors can be expressed via a form factor $F(\mathbf{q})$ as follows:
\begin{equation}
    \sum_{i, j=1}^{N_A}
    e^{ i \mathbf{q} \cdot \Delta \mathbf{r}_{ij} } \equiv
    N_A + \left( N_A^2 - N_A \right) |F({\mathbf q})|^2.
    \label{phase_factor_sum}
\end{equation}
In the limit of large atomic number density, the sum in the form factor can be approximated by an integral:
\begin{equation}
    |F({\mathbf q})|^2  = 
    \frac{1}{V^2}
    \int d^3 \mathbf{r}_i\,d^3 \mathbf{r}_j\,
    e^{ i \mathbf{q} \cdot \Delta {\mathbf r}_{ij} } .
    \label{formfactor}
\end{equation}
Therefore, the total differential cross section for DM scattering off a macroscopic target can be written as:
\begin{equation}
    \frac{d\sigma_\mathrm{tot}}{d {\mathbf q}} =
    A^2 
    \left[
        N_A + \left(N_A^2-N_A\right) |F({\mathbf q})|^2
    \right]
    \frac{d\sigma_{\chi N}}{d {\mathbf q}},
    \label{dsigmaA2}
\end{equation}
in the weak interaction regime. In contrast, in the strong interaction limit, the total scattering cross section saturates to the geometric size $S$ of the target\,\cite{Fukuda:2018omk, Fukuda:2021drn}.

Although Eq.~\eqref{acc_chi_N} provides a qualitative estimate, a full evaluation requires phase-space integration over the anisotropic DM velocity distribution in the laboratory frame, $f(\mathbf{v}_\chi)$, assuming the $z$-axis as the detection axis:
\begin{equation}
    \langle a_z \rangle =
    \frac{1}{m_\mathrm{tot}} 
    \frac{\rho_\chi}{m_\chi}
    \int d^3 \mathbf{v}_\chi\,d\Omega'_\chi\,
    q_z\,|\mathbf{v}_\chi|\,
    f(\mathbf{v}_\chi)\,
    \frac{d\sigma_\mathrm{tot} (\mathbf{q})}{d \Omega'_\chi}.
    \label{ave_a}
\end{equation}
Since the initial and final DM momenta share the same magnitude, $m_\chi |{\mathbf v}_\chi|$, the momentum transfer $\mathbf{q}$ is determined by the scattering angle of the outgoing DM, denoted $\Omega'_{\chi}$. Accordingly, the $\mathbf{q}$-dependence of the differential cross section can be expressed in terms of $\Omega'_\chi$. 


\section{EP-Testing Torsion Balance Exps} 

\begin{table*}[!t]
    \centering
    \setlength{\tabcolsep}{4pt}
    \renewcommand{\arraystretch}{1.4}
    \begin{tabular}{lrrrrrrc}
        \hline
        Experiment & Materials  & Mass & Shape & Structure & Outer Diameter & Height & $|\Delta a|$ \\
        \hline
        Roll et al. (1964)~\cite{Roll:1964rd}                           & Al/Au       & 30 g        & cylinder  & solid/solid  & 2.1\,cm/0.78\,cm   & 3.2\,cm    & $< 1.8 \cdot 10^{-11} \, \mathrm{cm/s^2}$ \\
        Braginskii et al. (1972)~\cite{Braginskii:1971tn}               & Al/Pt       & 0.49 g      & sphere    & solid/solid  & 0.70\,cm/0.35\,cm  & -          & $< 1.1 \cdot 10^{-12} \, \mathrm{cm/s^2}$ \\
        Eöt-Wash (1994)~\cite{Su:1994gu}                                & Be/Al       & 10 g        & cylinder  & solid/hollow & 2.0\,cm/2.0\,cm   & 1.7\,cm     & $ < 1.0 \cdot 10^{-11} \,                
                                                                                                    \mathrm{cm/s^2}$\footnotemark[1] \\
                                                                                                    & Be/Cu     & 10 g   & cylinder  & solid/hollow & 2.0\,cm/2.0\,cm   & 1.7\,cm   & $ < 1.0 \cdot 10^{-11} \,
                                                                                                    \mathrm{cm/s^2}$\footnotemark[1] \\
        Eöt-Wash (2008 \& 2012)~\cite{Schlamminger:2007ht,Wagner:2012ui} & Be/Al      & 5 g         & sphere\footnotemark[2]         & solid/hollow & 1.7\,cm/1.7\,cm   & -         & $ < 4.9 \cdot 10^{-13} \,                                                                                                 \mathrm{cm/s^2}$\footnotemark[1] \\[-4pt]
                                                                        & Be/Ti       & 5 g         & sphere\footnotemark[2]         & solid/hollow & 1.7\,cm/1.7\,cm   & -         & $ < 8.3 \cdot 10^{-13} \,                                 \mathrm{cm/s^2}$\footnotemark[1] \\
        \hline
    \end{tabular}
    \footnotetext[1]{The Eöt-Wash group typically reports acceleration limits along two orthogonal directions. In our analysis, we conservatively use the larger of the two.}
    \footnotetext[2]{The test bodies have irregular shapes and are modeled as spheres in our analysis.}
    \caption{Parameters of the test bodies and the \SI{95}{\percent} C.L.\ sensitivities to the acceleration difference $|\Delta a|$ between the test bodies in the four torsion-balance experiments.}
    \label{tab:acceleration_difference}
\end{table*}

\begin{figure*}[!t]
    \centering
    \includegraphics[width=0.4\textwidth]{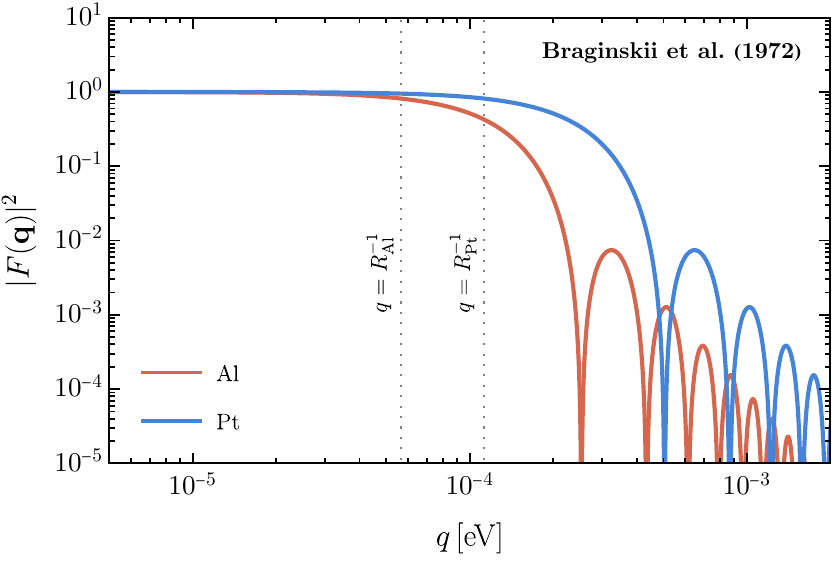}
    \hspace{20pt}
    \includegraphics[width=0.4\textwidth]{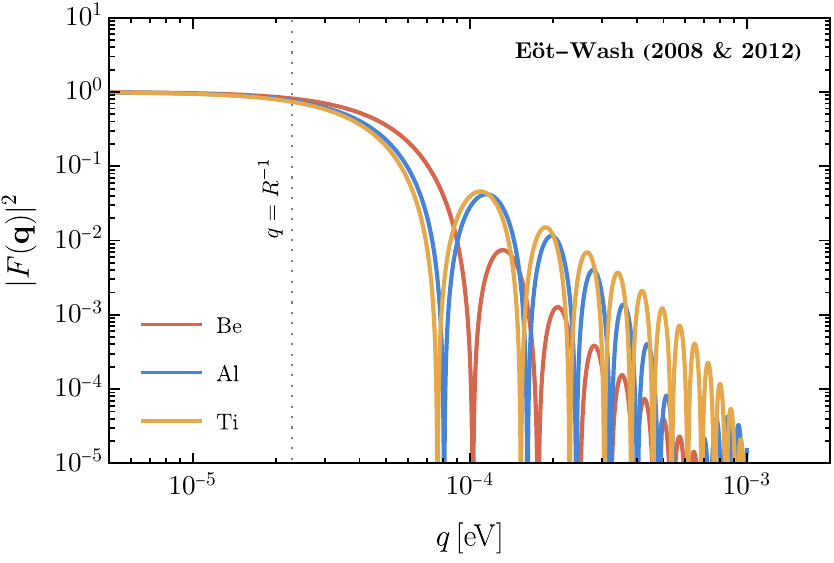}
    \caption{Form factors of the test bodies used in Braginskii et al.~\cite{Braginskii:1971tn} (left) and E\"ot-Wash (2008 \& 2012)~\cite{Wagner:2012ui} (right). In the former, the spherical radii are $R_\mathrm{Al}=0.35\,\mathrm{cm}$ and $R_\mathrm{Pt}=0.18\,\mathrm{cm}$; in the latter, the effective outer radius is $R=0.95\,\mathrm{cm}$.}
    \label{fig:form factor}
\end{figure*}

The acceleration in \geqn{ave_a} is, in principle, measurable with a force sensor. Among Earth-based experiments, torsion-balance setups provide some of the highest sensitivity to tiny accelerations: they suspend two sets of test bodies and measure the differential torque acting on them, i.e., the acceleration difference between the bodies. In EP tests, one typically compares test bodies with the same inertial mass; if the EP is violated, bodies made of different materials experience different solar gravitational forces, leading to an acceleration difference $\Delta a$.

DM scattering can also generate a differential acceleration of the test bodies, mimicking an EP-violating signal.
The reason is that although the test bodies are typically mass-matched, their differing material densities yield distinct volumes and spatial mass distributions. This asymmetry alters the scattering form factor $|F(\mathbf{q})|^2$ between DM and each test body\,\cite{Luo:2024ocg}, potentially generating measurable differential acceleration signals in practice,
\begin{equation}
    \Delta a \equiv \langle a^i_z \rangle - \langle a^j_z \rangle,
\label{signal}
\end{equation}
where $i$ and $j$ label the different test bodies. 
Therefore, the torsion-balance experiments designed to test the EP can also be used to search for DM scattering.

The null results from old torsion-balance experiments in EP testing can place limits on DM interaction. We focus on four benchmark experiments. The first two, by Roll et al.~\cite{Roll:1964rd} and Braginskii et al.~\cite{Braginskii:1971tn}, employed solid test bodies of different materials and sizes. They were designed to search for EP-violating forces exerted by the Sun, which vary periodically due to Earth's rotation. The later Eöt-Wash experiments~\cite{Su:1994gu,Schlamminger:2007ht,Wagner:2012ui} used self-rotating torsion balances with an optional rotation period (like 2\,h) to effectively suppress or distinguish the low-frequency noise and diurnal backgrounds. Due to Earth's rotation, the local gravitational vector has a small horizontal component. As a result, Earth acted as an attractor, exerting EP-violating forces on the test bodies; their horizontal components induced a measurable torque whose frequency is determined by the rotation period. To reduce systematics, the Eöt-Wash experiments used test bodies with identical external shapes but different internal structures: the Be test masses were solid, whereas those made of heavier materials (Al, Cu, Ti) were hollow. The shell thickness of the hollow test bodies are determined from their densities and outer dimensions. It should also be noted that Eöt-Wash (2008 \& 2012)~\cite{Schlamminger:2007ht,Wagner:2012ui} employed irregularly shaped test bodies, which are approximated as spheres in our analysis for simplicity. These test bodies and the sensitivities of the four experiments are summarized in Table~\ref{tab:acceleration_difference}.
The Eöt-Wash group also performed non-rotating torsion-balance searches for ultralight forces from a linearly coupled $B\!-\!L$ gauge boson~\cite{Shaw:2021gnp}. However, that analysis targets signals in the $\sim$\,mHz band, whereas in a non-rotating torsion balance the DM-scattering signal would appear at the sidereal frequency. Owing to the mismatch in signal frequencies, the analysis here is not suitable for constraining our DM-scattering scenario.

The test bodies used in the above experiments can be classified into four geometries: solid spheres, hollow spheres, solid cylinders, and hollow cylinders. To evaluate the coherent DM--test-body scattering cross sections for these geometries, we need the corresponding form factors. For a solid sphere of radius $R$, the form factor is obtained by directly integrating Eq.~\eqref{formfactor} as follows:
\begin{equation}
    F_S(\mathbf{q}) = \frac{3}{(qR)^3} \left[ \sin(qR) - qR \cos(qR) \right] .
    \label{eq:FF_sphere}
\end{equation}
For a solid cylinder of radius $R$ and height $h$, the form factor depends on the momentum transfer components parallel ($q_h$) and perpendicular ($q_R$) to the cylinder axis,
\begin{equation}
    F_C(\mathbf{q}) 
    = 
    \frac{1- e^{i q_h h}}{i q_h h}
    \frac{2 J_1(q_R R)}{q_R R} .
    \label{eq:FF_cylinder}
\end{equation}
where $J_1(x)$ is the Bessel function of the first kind. A hollow spherical shell or hollow cylinder can be treated as a larger solid shape with a smaller one removed from the center. The form factors for a hollow sphere, $F_{HS}(\mathbf{q})$ and hollow cylinder, $F_{HC}(\mathbf{q})$, can then be derived from those of the corresponding solid shapes as follows:
\begin{equation}
    F_{HS,HC} (\mathbf{q})
    = 
    \frac{
        V_\mathrm{out} F^\mathrm{out}_{S,C} (\mathbf{q})
        - 
        V_\mathrm{in} F^\mathrm{in}_{S,C} (\mathbf{q})
    }{ V_\mathrm{out} - V_\mathrm{in} } .
    \label{eq:FF_hollow}
\end{equation}
Here, the labels \textit{out} and \textit{in} refer to the outer and inner dimensions of the test bodies, indicating the corresponding volumes $V$ and form factors $F$.
All form factors approach $1$ in the small-$\mathbf{q}$ limit and vanish in the large-$\mathbf{q}$ limit.

The form factors of the test bodies used in the Braginskii et al.\ (1972) and E\"ot-Wash (2008 \& 2012) experiments are shown in Fig.~\ref{fig:form factor}. Coherent effects are suppressed when the scattering length becomes shorter than the target size, $\lambda \lesssim R$, as expected. In the former experiment, the different radii of the Al and Pt spheres lead to notable deviations in their form factors for $q \gtrsim R_{\rm Al}^{-1}$. In the latter, where the test bodies have similar outer radii, the form factors differ only at large $q$ that resolve the internal hollowness, and the difference is much smaller than in the case with distinct outer dimensions. 
As the DM wavelength increases and $q$ decreases, the form factors approach unity, so the form-factor difference shrinks and the resulting differential acceleration becomes smaller. Consequently, the torsion-balance sensitivity would be reduced.


\section{Constraints} 


When DM scattering is isotropic, $d\sigma_{\chi N}/d\Omega'_\chi=\sigma_{\chi N}/(4\pi)$, the differences in the scattering cross sections among the test bodies in Eq.~\eqref{ave_a} arise solely from their form factors. These form-factor differences induce a differential acceleration via coherent DM scattering in EP-testing experiments. The null results of these searches for EP-violating forces therefore set upper limits on the acceleration difference $|\Delta a|$ between the test bodies (see Table~\ref{tab:acceleration_difference}). A DM-induced scattering force would manifest in the same way as an EP-violating force at the same modulation frequency. Hence, the predicted difference between the $i$th and $j$th test bodies, $\Delta a=\langle a^i_z\rangle-\langle a^j_z\rangle$, must not exceed the experimental bound on $|\Delta a|$. This requirement in turn constrains the cross section $\sigma_{\chi N}$, namely the strength of DM interactions with nucleons.

\begin{figure}[!t]
    \centering
    \includegraphics[width=0.45\textwidth]{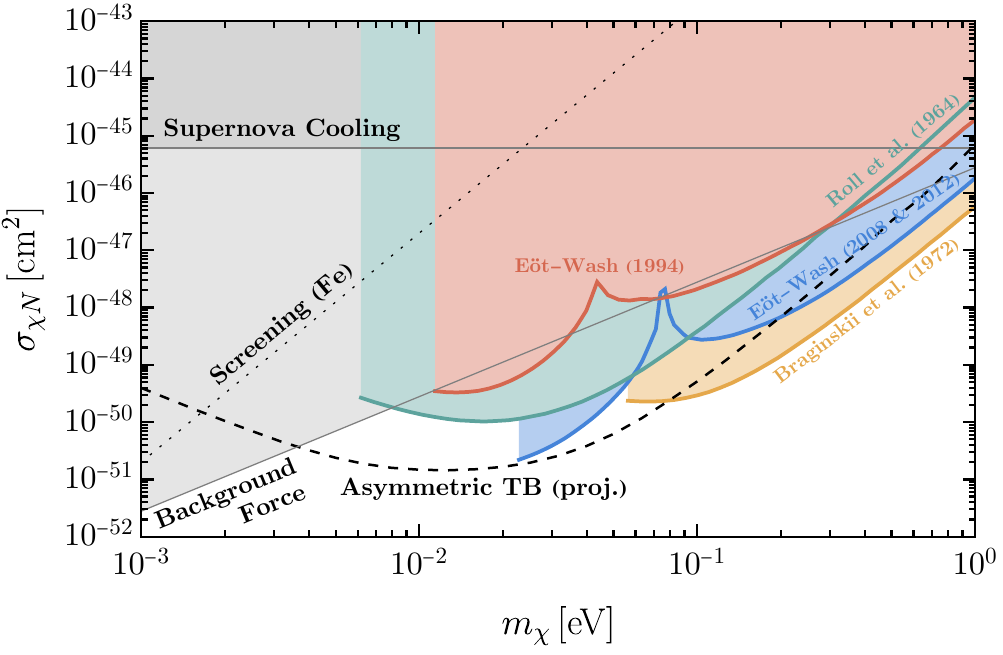}
    \caption{Constraints on the DM--nucleon cross section from existing EP tests (colored solid lines). The dashed black curve shows the projected sensitivity of an asymmetric rotating torsion balance~\cite{Luo:2024ocg}. The gray shaded regions are excluded by supernova cooling and DM-induced background forces. The dotted curve indicates possible screening by a thick iron wall.}
    \label{fig:constraint}
\end{figure}

The resulting limits from the above experiments are shown in Fig.~\ref{fig:constraint}. The strongest sensitivity is attained when the DM de Broglie wavelength is comparable to the test-body size. For heavier DM, coherence is lost and the constraints weaken. Around $m_\chi \sim 5\times10^{-2}\,\mathrm{eV}$, the sensitivities of E\"ot-Wash (1994) and E\"ot-Wash (2008 \& 2012) decrease because their test bodies share identical outer geometries, giving nearly identical form factors. As seen in Fig.~\ref{fig:form factor}, the red and orange curves intersect at $q \sim 9\times10^{-5}\,\mathrm{eV}$, making the corresponding accelerations almost equal and suppressing the differential signal $\Delta a$. Other crossings are averaged out in the momentum integration of Eq.~\eqref{ave_a} and therefore do not lead to comparable reductions in sensitivity.
For lighter DM, the wavelength can exceed the size of the experimental apparatus. In this case, scattered waves from the test body and from surrounding structures can interfere, so the notion of an individual-body cross section becomes ill-defined and the present analysis no longer applies. We therefore impose a cutoff at the DM mass for which the DM wavelength equals the outer dimension of the test bodies.

We find that torsion-balance experiments provide the most stringent constraints to date on quadratically coupled sub-eV DM, thereby complementing searches for other types of sub-eV DM (i.e., linearly coupled candidates) such as axions~\cite{Sikivie:2020zpn}, dark photons~\cite{Caputo:2021eaa}, chameleons~\cite{Li:2016tux,Elder:2016yxm}, and symmetrons~\cite{Cronenberg:2018qxf}.

If the DM interaction is sufficiently strong, screening effects inside matter can become significant~\cite{Hees:2018fpg, Banerjee:2022sqg, Day:2023mkb}. Given the DM--nucleon coupling $\mathcal{L} = - \chi^2 \bar{N} N / (2f)$, DM acquires an effective mass in materials. If this effective mass is large, the DM wind is exponentially suppressed within matter and thus screened. The region in which a sufficiently thick iron wall screens the DM wind is indicated by the dotted curve in \gfig{fig:constraint}. However, this constraint does not apply if the DM--nucleon potential is attractive. See Appendix~A for further details.

DM within the same mass range is subject to other indirect constraints. First, the same coupling leads to DM production in supernovae via nucleon bremsstrahlung, $N + N \to N + N + \chi + \chi$, thereby affecting supernova cooling. Observations require that the emission of light DM not significantly exceed that of neutrinos~\cite{Raffelt:1996wa}. The corresponding constraints~\cite{Olive:2007aj} are shown as the shaded gray region in \gfig{fig:constraint}. Additionally, quadratically coupled DM can induce forces between nucleons~\cite{VanTilburg:2024xib, Barbosa:2024pkl, Grossman:2025cov, Cheng:2025fak, Gan:2025nlu}. When a DM wave scatters off a source nucleon, it modifies the effective mass of a nearby test nucleon, which then experiences a force proportional to the gradient of this mass. The resulting constraint~\cite{VanTilburg:2024xib} from short-range EP tests~\cite{Tan:2020vpf} appears as the light gray shaded region in \gfig{fig:constraint}. Moreover, quadratically coupled DM can be upscattered by cosmic rays~\cite{Cappiello:2018hsu, Bringmann:2018cvk}, enabling its detection in conventional direct-detection and neutrino experiments. Current constraints on the scattering cross section reach $\sigma \lesssim 4 \cdot 10^{-33} \, \mathrm{cm^2}$~\cite{PandaX-II:2021kai, Super-Kamiokande:2022ncz, LZ:2025iaw}, whereas our results strengthen these upper bounds by more than seventeenth orders of magnitude.

\section{Discussion and Conclusions} 

We have shown that although detecting single scattering events from light DM is challenging due to limited energy transfer, the acceleration of macroscopic objects caused by continuous scattering offers a novel observable. Coherent scattering significantly enhances both the cross section and the resulting acceleration signal. Such effects are measurable using torsion balance experiments with asymmetric configurations. We systematically analyzed several state-of-the-art torsion balance setups originally developed for weak equivalence principle tests, deriving constraints on DM-nucleon scattering across the mass range $(10^{-2},1)\,$eV. Compared with existing bounds, these experiments exhibit superior sensitivity. 



Modern high-sensitivity rotating torsion-balance experiments typically employ test bodies with identical outer dimensions to minimize systematic effects. However, as shown in Fig.~\ref{fig:form factor}, the form-factor difference between such test bodies is much smaller than in the case with different outer dimensions. We therefore propose a rotating torsion balance using test bodies with different outer dimensions, accepting a controllable increase in backgrounds. This configuration can further enhance the sensitivity to coherent DM scattering, and the projected reach is shown by the dashed line in Fig.~\ref{fig:constraint}.
This opens a new avenue for direct detection of low-mass DM.


\section*{Acknowledgements}

\noindent
The authors thank Peter Cox, Shao-Feng Ge, Peng-Shun Luo, and Yevgeny Stadnik for fruitful discussions.
S.~M.\ was supported by the Grant-in-Aid for Scientific Research from the Ministry of Education, Culture, Sports, Science and Technology, Japan (MEXT), under Grant No.\ 24H00244 and 24H02244. 
J. S. is supported by the the Japan Society for the Promotion of Science (JSPS) as a part of the JSPS Postdoctoral Program (Standard) with grant number: P25018, National Natural Science Foundation of China (Nos. 12375101, 12425506, 12090060, 12090064), the SJTU Double First Class start-up fund WF220442604.
S.~M. and J. S. are also supported by the World Premier International Research Center Initiative (WPI), MEXT, Japan (Kavli IPMU)
C.-Y. Xing is supported by the Fundamental Research Funds for the Central Universities (No.~24CX06048A).  
L. Zhu is supported by the National Natural Science Foundation of China (No. 12005308 and No. 12150012).

\begin{appendix}
\section{Appendix A -- Screening Effects}
\label{appA}

\noindent
The coupling between DM ($\phi$) and nucleons is given by
\begin{equation}
    \mathcal{L} = \pm \frac{1}{2f} \phi^2 \bar{N} N,
\end{equation}
where the positive (negative) sign corresponds to an attractive (repulsive) potential between DM and nucleons. Here, we discuss in detail how such interactions influence the screening effects of DM in realistic experiments.

\vspace{0.2cm}

\noindent
\textbf{Repulsive Potential} -- The interaction Lagrangian, $\mathcal{L} = - \phi^2 \bar{N} N / 2f$, effectively behaves as a DM mass term when the nucleon field acquires an expectation value $\braket{\bar{N} N} = n_N$ in matter, with $n_N$ denoting the nucleon number density. Then, DM acquires an effective mass,
\begin{equation}
    \Delta m_\phi^2 = \frac{1}{f} \langle \bar{N} N \rangle = \frac{n_N}{f},
\end{equation}

When DM with mass $m_\chi$ and momentum ${\mathbf p}_\chi$ enters a material medium from vacuum, the dispersion relation
\begin{equation}
    \sqrt{ \mathbf{p}_\chi^{\prime2} + m_\chi^2 + \Delta m_\phi^2 } 
    = 
    \sqrt{ \mathbf{p}_\chi^2 + m_\chi^2 },
\end{equation}
dictates that the DM momentum inside matter $\mathbf{p}'_\chi$ is, 
\begin{equation}
    | \mathbf{p}'_\chi | =  \sqrt{ \mathbf{p}_\chi^2 - \Delta m_\phi^2 }.
\end{equation}
If the coupling is strong enough that the DM effective mass exceeds its vacuum momentum, $\Delta m_\phi > |{\mathbf p}_\chi|$, the DM momentum inside the material becomes imaginary. This implies that the DM wave is exponentially attenuated while passing through the material. When the material thickness exceeds the DM penetration depth, the wave is fully reflected, resulting in a \textit{screening effect}. Only when $\Delta m_\phi < |{\mathbf p}_\chi|$ can DM propagate through the material. In the main text (and in Fig.\,\ref{fig:constraint}), we adopt this as the condition for DM to remain unshielded.

For the experiments under consideration, potential sources of DM shielding include the Earth's atmosphere and the external shield of the torsion balance apparatus. The mass density of the atmosphere near sea level is observed to be $\rho_\mathrm{atmo} \sim 0.0012 \, \mathrm{g}/\mathrm{cm}^3$, corresponding to a nuclear number density of $n_\mathrm{atmo} \simeq \rho_\mathrm{atmo}/1\,\text{GeV}$.
To ensure that DM is not significantly attenuated in the atmosphere, the following condition must be satisfied:
\begin{equation}
    f \gtrsim \frac{n_\mathrm{atmo}}{m_\phi^2 v_\phi^2} \sim 5 \cdot 10^{7} \times \left( \frac{m_\phi}{10^{-2}\,\mathrm{eV}} \right)^{-2} \, \mathrm{GeV} .
\end{equation}
Translated to the scattering cross-section, this becomes,
\begin{equation}
    \sigma_{\chi N} \lesssim  6 \cdot 10^{-40} \times \left( \frac{m_\phi}{10^{-2}\,\mathrm{eV}} \right)^{4} \, \mathrm{cm}^2 .
\end{equation}

Assuming the torsion balance apparatus is enclosed in a thick iron layer with density $\rho_\mathrm{Fe} \sim 7.87 \, \mathrm{g}/\mathrm{cm}^3$ (other materials would not significantly affect the result), the condition for DM to penetrate the shield is:
\begin{equation}
    \sigma_{\chi N} \lesssim  2 \cdot 10^{-47} \times \left( \frac{m_\phi}{10^{-2}\,\mathrm{eV}} \right)^{4} \, \mathrm{cm}^2 .
    \label{shield_iron}
\end{equation}
Due to iron's higher nuclear number density, it imposes stronger constraints on the cross-section. In \gfig{fig:constraint}, the condition in \geqn{shield_iron} is shown as a black dotted line. For DM-nucleon interactions with a repulsive potential, the parameter space above this line may be shielded.

Reference~\cite{Day:2023mkb} describes the modification of the DM effective mass due to interactions as a Meissner-like effect. 
Essentially, this is analogous to the quantum mechanical scattering of a particle by a potential barrier: 
when the barrier height exceeds the particle’s kinetic energy, the transmission amplitude is exponentially suppressed.

\vspace{0.2cm}

\noindent
\textbf{Attractive Potential} -- The positive interacting Lagrangian, $\mathcal{L} = + \phi^2 \bar{N} N / (2f)$, represents an attractive potential between DM and nucleons. It induces a negative effective mass-squared term for the DM field, $\Delta m_\phi^2 = - n_N / f$. In this case, the dispersion relation becomes
\begin{equation}
    \sqrt{ \mathbf{p}_\chi^{\prime2} + m_\chi^2 - |\Delta m_\phi^2|} 
    = 
    \sqrt{ \mathbf{p}_\chi^2 + m_\chi^2 },
\end{equation}
dictating that the DM momentum inside matter,
\begin{equation}
    | \mathbf{p}'_\chi | =  \sqrt{ \mathbf{p}_\chi^2 + |\Delta m_\phi^2| },
\end{equation}
is always positive. This implies that DM can still propagate inside matter. Consequently, when DM--nucleon interactions are attractive, no screening effect arises, and the constraints shown in \gfig{fig:constraint} become irrelevant.

\end{appendix}

\bibliographystyle{utphys}
\bibliography{ref}

\end{document}